\documentclass[reprint,
superscriptaddress,
amsmath,amssymb,
aps,prd,floatfix,
twocolumn,
showpacs,
]{revtex4-2}

\usepackage{amsthm}
\usepackage{tikz}
\usepackage{graphicx}
\usepackage{dcolumn}
\usepackage{bm}
\usepackage{braket}
\usepackage[english]{babel}
\usepackage{hyperref}
\usepackage{amsmath}
\usepackage{graphicx}
\usepackage{float}
\usepackage{siunitx}
\usepackage{xcolor}
\usepackage{xspace}
\usepackage{eucal}
\usepackage{ulem}
\usepackage{multirow}
\usepackage[utf8]{inputenc}
\usepackage{pgfplots}
\pgfplotsset{compat=1.14}
\usepackage{upgreek}
\usepackage{siunitx}
\usepackage{gensymb}
\usepackage{amsmath}
\usepackage[justification=justified,singlelinecheck=false]{caption}
\usepackage{subcaption}
\usepackage{textcomp}
\usepackage{comment}
\usepackage{cleveref}

\Crefname{figure}{Figure}{Figures}

\usepackage{todonotes}
\usepackage{tabularx}
\usepackage{booktabs}
\usepackage{comment}
\usepackage{array}
\DeclareSIUnit\g{g}
\DeclareSIUnit\gal{Gal}
\DeclareSIUnit\torr{Torr}
\DeclareSIUnit\bar{Bar}
\DeclareSIUnit\kelvin{K}
\DeclareSIUnit\inch{inch}
\DeclareSIUnit\joule{J}
\DeclareSIUnit\rad{rad}

\begin{document}

\title{Optical Truss Interferometer for the LISA Telescope}

\author{Kylan Jersey}
\affiliation{Texas A\&M University, Aerospace Engineering \& Physics, 701 H.R. Bright Bldg, College Station, TX~77843, USA}
\affiliation{Wyant College of Optical Sciences, The University of Arizona, 1630 E. University Blvd, Tucson, AZ~85719, USA}
\author{Ian Harley-Trochimczyk}
\affiliation{Texas A\&M University, Aerospace Engineering \& Physics, 701 H.R. Bright Bldg, College Station, TX~77843, USA}
\author{Yanqi Zhang}
\affiliation{Texas A\&M University, Aerospace Engineering \& Physics, 701 H.R. Bright Bldg, College Station, TX~77843, USA}
\affiliation{Wyant College of Optical Sciences, The University of Arizona, 1630 E. University Blvd, Tucson, AZ~85719, USA}
\author{Felipe Guzman}\email[Electronic mail: ]{felipe@tamu.edu}
\affiliation{Texas A\&M University, Aerospace Engineering \& Physics, 701 H.R. Bright Bldg, College Station, TX~77843, USA}


\begin{abstract}
The LISA telescopes must exhibit an optical path length stability of $\frac{\mathrm{pm}}{\sqrt{\mathrm{Hz}}}$ in the mHz observation band to meet mission requirements. The optical truss interferometer is a proposed method to aid in the ground testing of the telescopes, as well as a risk-mitigation plan for the flight units. This consists of three Fabry-Perot cavities mounted to the telescope which are used to monitor structural displacements. We have designed and developed a fiber-based cavity injection system that integrates fiber components, mode-matching optics, and a cavity input mirror into a compact input stage. The input stages, paired with return mirror stages, can be mounted to the telescope to form the optical truss cavities. We performed a thorough sensitivity analysis using various simulation methods to support the fabrication and assembly of three first-generation prototype cavities, each of which exhibited a satisfactory performance based on our models.
\bigbreak

\copyright 2023 Optica Publishing Group under the terms of the Open Access Publishing Agreement. Users may use, reuse, and build upon the article, or use the article for text or data mining, so long as such uses are for noncommercial purposes and appropriate attribution is maintained. All other rights are reserved.
\end{abstract}

\maketitle
\section{Introduction}
\label{sec:Intro}
The Laser Interferometer Space Antenna (LISA) \cite{wanner_2019,thorpe_2010} will be the first space-borne gravitational wave observatory meant for the detection of gravitational waves emitted by low-frequency sources in the observation band between 0.1 mHz and 1 Hz. This mission is led by the European Space Agency (ESA), with contributions from the National Aeronautics and Space Administration (NASA) and an international consortium of scientists, to fly a constellation of three spacecraft, each separated by 2.5 million kilometers to form an equilateral triangle in their formation. The spacecraft will relay laser beams between each other to form a set of long baseline laser interferometers whose endpoints are free flying test masses housed by the spacecraft. The primary interferometric goal is to measure variations in the separation between the free flying test masses onboard different spacecraft with a sensitivity of around $10 \frac{\mathrm{pm}}{\sqrt{\mathrm{Hz}}}$ level at mHz frequencies\cite{jennrich_2009}. The LISA telescopes are bidirectional optical systems used to expand and send the outgoing beams to be captured by the other spacecraft while also capturing a small fraction of large 15 km-wide incoming beams from the other spacecraft.
Since these telescopes lie directly in the optical path of the long arm interferometers, their structure must be dimensionally stable at the $\frac{\mathrm{pm}}{\sqrt{\mathrm{Hz}}}$ level within the mHz frequency band to allow for the proper detection of incoming gravitational wave signals, and as such, they are to be constructed with highly stable low-expansion materials. During ground testing, telescope prototypes must be measured and verified to meet the stability requirements. While the primary measurement is to probe the overall path length stability along the optical axis of the telescope, there is also a motivation to measure the length stability at multiple locations around the telescope aperture to reconstruct wavefront errors introduced by structural distortions. This can be done with an optical truss interferometer (OTI), a set of three Fabry-Perot cavities that can be mounted around the telescope to monitor structural distortions over time.

\begin{figure}[h]
\centering
\includegraphics[width=\linewidth]{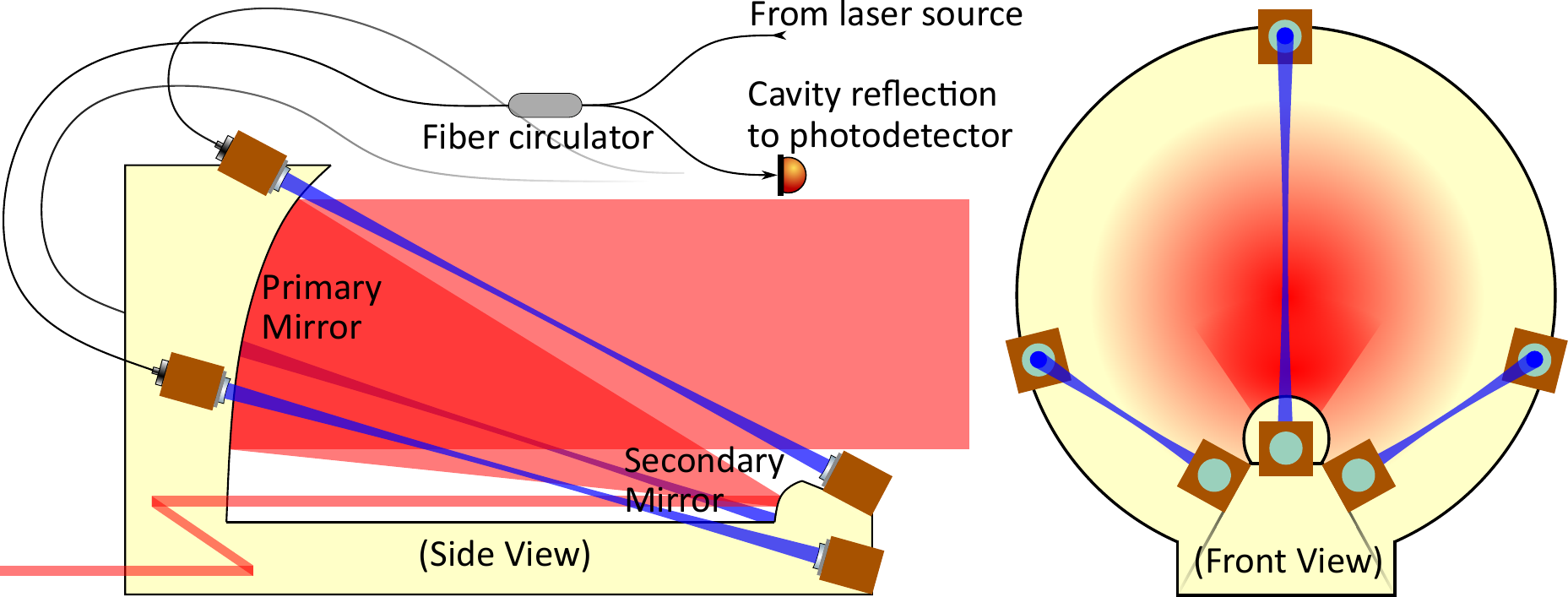}
\caption{Graphical drawing of the truss cavities mounted onto the LISA telescope structure. The fiber-injected input stages are attached around the primary mirror, while the return stages are attached around the secondary mirror.}
\label{OTI_inkscape}
\end{figure}

The OTI system depicted in Figure \ref{OTI_inkscape} is composed of three optical truss cavities, each mounted lengthwise at different lateral positions around the telescope. We have designed and developed compact fiber-based units that integrate single-mode polarization-maintaining (PM) fiber, mode matching optics, and a cavity input mirror into a modular input stage for each OTI cavity\cite{OTI_SPIE}. These input stages can be mounted around the primary mirror of the telescope by means of hydroxide catalysis bonding\cite{robertson_2013}, while the return stages, which house the cavity return mirrors, can be attached in a similar manner around the secondary mirror of the telescope. Utilizing a Pound-Drever-Hall (PDH) \cite{black_2001} frequency locking scheme, displacements in each cavity along the optical axis will cause proportional variations in the frequency of light emitted from the corresponding 1064 nm laser. Ideally, variations in the cavity lengths will be solely due to displacements in the telescope structure along the three different lateral positions. Thus, monitoring the laser frequencies stabilized to each OTI cavity will effectively monitor the displacement noise in the structure on which the cavities are mounted. The optical truss interferometer serves as a risk-mitigation plan to aid in the verification of $\frac{\mathrm{pm}}{\sqrt{\mathrm{Hz}}}$ stability in the telescope prototypes during ground testing and, if necessary, can be used to monitor the telescopes during flight. This paper will cover the optomechanical design, fabrication, and assembly of a first-generation OTI prototype.


\section{Optomechanical Design \& Sensitivity Analysis}
\label{sec:design}
\vspace{-0.2cm}
\subsection{Cavity Geometry}
\label{sec:2.1}
\vspace{-0.2cm}

\begin{figure*}[htpb]
    \centering
    \includegraphics[width=8.9cm]{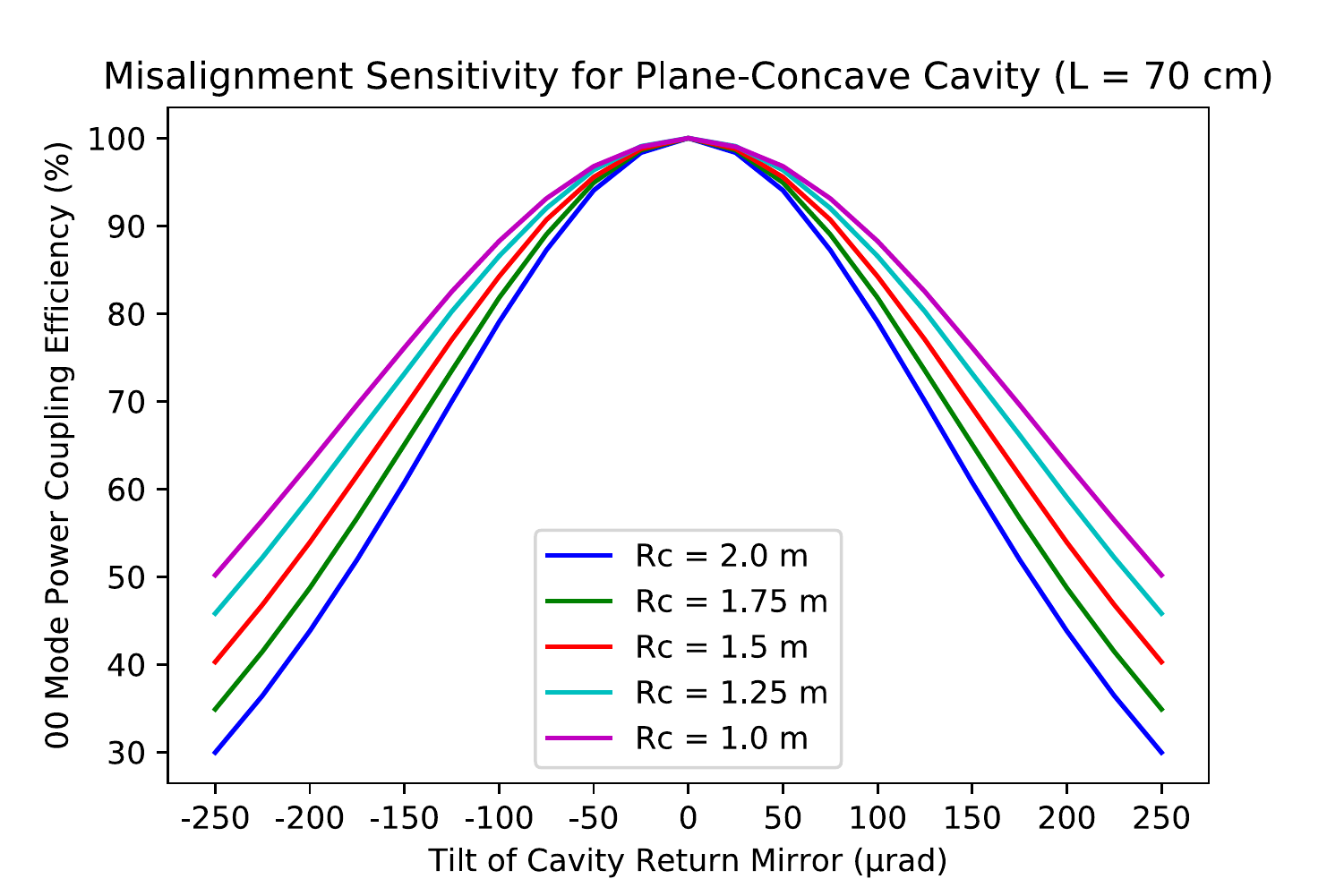}
    \includegraphics[width=8.9cm]{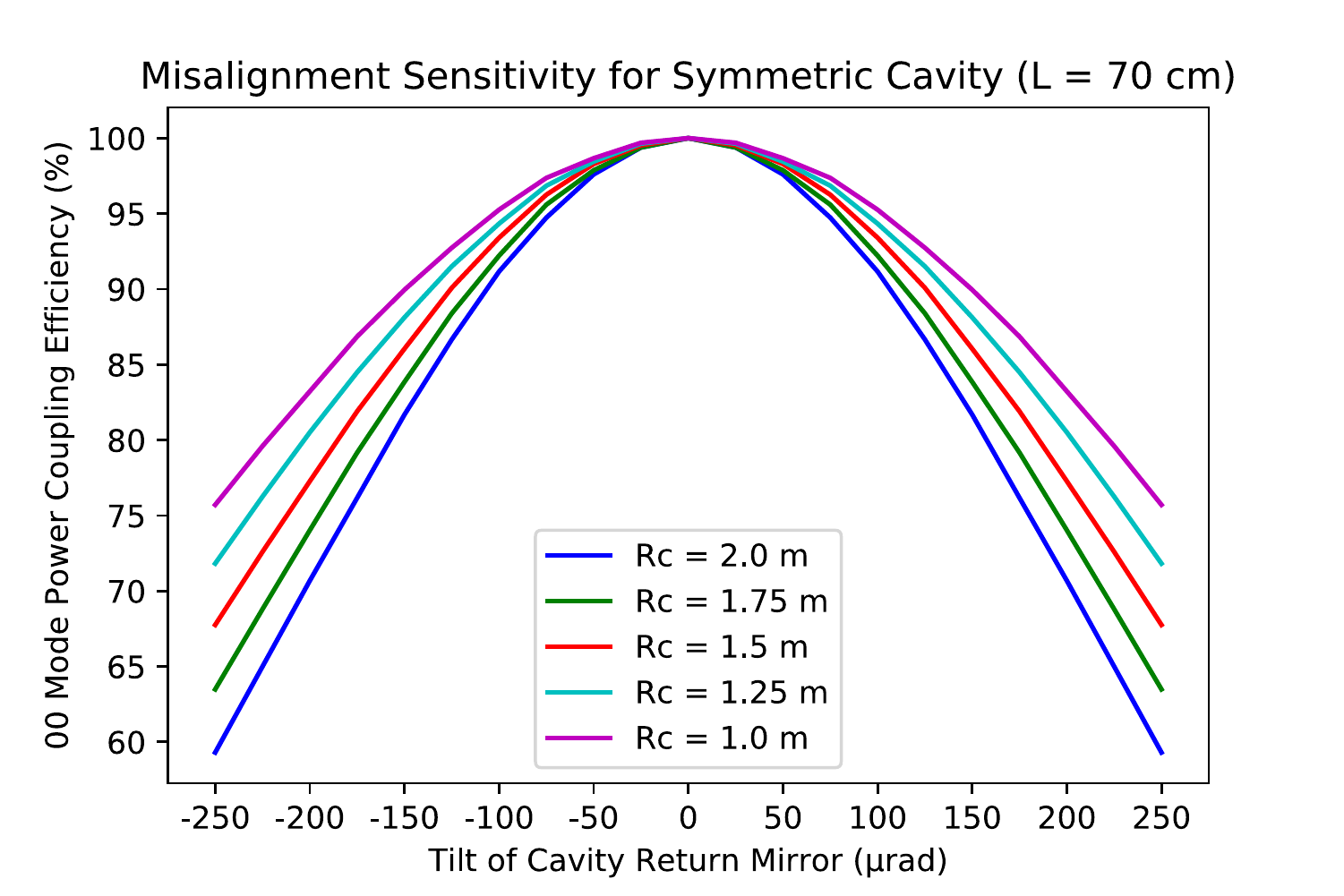}
    \caption{Early simulation results from using FINESSE to investigate the misalignment sensitivity for plane-concave (left) and symmetric (right) cavity geometries with various mirror curvatures. In both cases, the cavity return mirror is tilted using a standard mirror misalignment feature in FINESSE \cite{brown_freise_2020}.}
    \label{Cavity Geometry}
\end{figure*}

The optical design work for the OTI initially involved an ideal cavity geometry design along with a compact mode matching system to accompany the input stage. The choice of cavity geometry was the first concern in the design process since the components needed for the input stage depend upon both the cavity length and the curvature of the cavity mirrors. While the overall cavity length is ultimately determined by the dimensions of the LISA telescope, a baseline length of 70 cm was given as a starting point. The curvature of the cavity mirrors must allow for a stable cavity to occur, and the stability condition \cite{siegman_1986} for this is shown by Equation \ref{Cavity Stability}

\begin{equation}
\label{Cavity Stability}
    0 \le g_{1}g_{2} = \left(1-\frac{L}{R_1}\right)\left(1-\frac{L}{R_2}\right) \le 1
\end{equation}

\noindent where $L$ is the cavity length, and $R_1$ and $R_2$ are the radii of curvature of the first and second cavity mirrors, respectively. While there are multiple solutions that satisfy the stability condition above, the most fitting cavity geometry for the OTI is that which is most robust to misalignment of the cavity mirrors since there are not many intrinsic degrees of freedom to aid in the alignment of the cavity once the OTI units are assembled and mounted onto a testing structure. We conducted various simulations using the software package FINESSE \cite{brown_freise_2020} to investigate the effect of mirror misalignment on the power coupling efficiency of the fundamental transverse cavity mode, commonly known as the 00 mode, for various cavity geometries. This power coupling efficiency is defined as $V = \frac{P_{max}-P_{min}}{P_{max}+P_{min}}$ when measuring the optical power reflected from the cavity upon the laser frequency scanning over the 00 mode resonance, where $P_{max}$ is the power measured off-resonance and $P_{min}$ is the power measured exactly on resonance. Our earliest investigations on plane-concave and symmetric cavity geometries using FINESSE simulations have shown that the symmetric cavity geometry has a lower sensitivity to mirror misalignment than the plane-concave geometry, as shown in Figure \ref{Cavity Geometry}. Furthermore, we found that the misalignment sensitivity is dependent on the radius of curvature of the mirror(s). Specifically, the power coupling to the 00 mode will become less sensitive to mirror misalignment as the radius of curvature of the mirror(s) is decreased \cite{OTI_SPIE}. However, as the cavity geometry approaches the limit of the stability condition in Equation \ref{Cavity Stability}, the system becomes more sensitive to mirror misalignment \cite{siegman_1986}. Thus, we chose a symmetric cavity whose mirrors have a radius of curvature of 500 mm, giving a relatively low misalignment sensitivity while simultaneously allowing for some freedom in the overall cavity length.

\subsection{OTI Input Stages}
\label{sec:2.2}
\begin{figure*}[htpb]
    \centering
    \includegraphics[width=\linewidth]{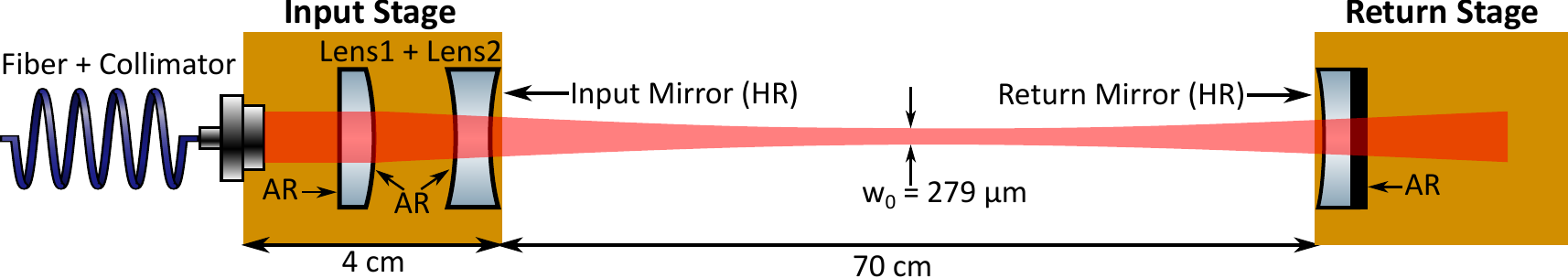}
    \caption{Schematic of the optical design of an OTI cavity. The mode matching lenses take up $\approxeq$ 3 cm in length and the cavity input mirror is coated onto the back surface of the second lens. The high-reflectance (HR) mirror surfaces have been coated such that $R \ge 0.998$ at 1064 nm, giving the cavity a finesse of $\gtrapprox$ 1600.}
    \label{Inputstage}
\end{figure*}

Once an ideal cavity geometry was established, we focused on designing the OTI input stage which integrates fiber injection, mode matching optics, and the cavity input mirror into a single compact unit. We designed a mode matching configuration that transforms the transverse profile of an incident beam emitted from a fiber collimator into the cavity eigenmode. This included a specific combination of a fiber, collimator, and mode matching lenses (Figure \ref{Inputstage}). Furthermore, this optical system was designed for a small form factor such that these components can be seated in a housing of just a few centimeters in size.

The beam emitted from the fiber collimator shown in Figure \ref{Inputstage} is approximately a Gaussian beam with a waist size of 0.95 mm and is transformed by the two mode matching lenses into a beam with a 279 $\mu$m waist located at the center of the cavity. The first lens is a simple plane-convex positive lens with anti-reflective (AR) coating, while the second lens is a multi-purpose component comprised of a concave AR-coated surface with negative refractive power followed by the cavity input mirror formed by a dielectric coating ($R \ge 99.8\%$) on the back surface of the lens. The mode matching calculations are simple enough for an analytical solution, but for purposes of further simulation and modeling we used multiple software packages, JamMT \cite{jammt} and Zemax OpticStudio \cite{zemax}, to optimize the optical system. JamMT is a Java application that can be used to determine mode matching configurations based on the transverse cavity eigenmode, the incident mode, and a selection of lenses. Once an approximately fitting optical system was determined, we then recreated the system in a Zemax environment to optimize the parameters to produce the desired transverse cavity mode as well as to maintain a small form factor for the input stage (see Table \ref{Optical Design Table}). This optical system allows for the cavity input and return stages to be housed and secured in handheld units that can be mounted to a telescope or testing structure.

\begin{table}[htpb]
    \centering
\begin{tabular}{|m{1.9cm}||m{1.9cm}|m{1.9cm}|m{1.9cm}|}
\hline
 \textbf{Fiber} & \textbf{Wavelength} & \textbf{Mode Field Diameter} & \textbf{Fiber NA} \\
\hline \hline
 single-mode PM cable & 1064 nm & $6.6 \pm 1 \mu$m & 0.12 \\
\hline \hline
 \textbf{Collimator} & \textbf{Focal Length} & \textbf{AR Coating} & \textbf{Output Mode} \\
\hline
 adjustable mounting & 12 mm & R$ < 0.4\% $ & $\omega_{0} = 0.95$ mm \\
\hline \hline
 \textbf{Lens 1} & \textbf{Focal Length} & \textbf{Position} & \textbf{Coatings} \\
\hline
 \textbf{Surface 1:} plane & $\infty$ & incident beam waist & both sides \\
 \textbf{Surface 2:} convex & 35.5 mm & 4.45 mm (thickness) & R$ < 0.3\%$ \\
\hline \hline
 \textbf{Lens 2} & \textbf{Focal Length} & \textbf{Position} & \textbf{Coatings} \\
\hline
 \textbf{Surface 1:} concave & -19.3 mm & 17.26 mm (from lens 1) & R$ < 0.3\%$ \\
 \textbf{Surface 2:} concave & -1112.1 mm & 5.57 mm (thickness) & R $ \ge 99.8\%$ \\
\hline
\end{tabular}
\caption{Optical design parameters for the OTI input stage. The fiber cable and collimator used here are commercially available products, and their specifications were assumed to be constant during optimization. The focal length, thickness, and position of each fused-silica lens were the variables used to optimize the mode-matching system. Note that Surface 2 of Lens 2 is the cavity input mirror, and thus was not a variable in optimization and does not significantly contribute to the focusing power of Lens 2.}
\label{Optical Design Table}
\end{table}

\subsection{Sensitivity Analysis}
\label{sec:2.3}

\begin{figure*}[htpb]
    \centering
    \includegraphics[width=8.9cm]{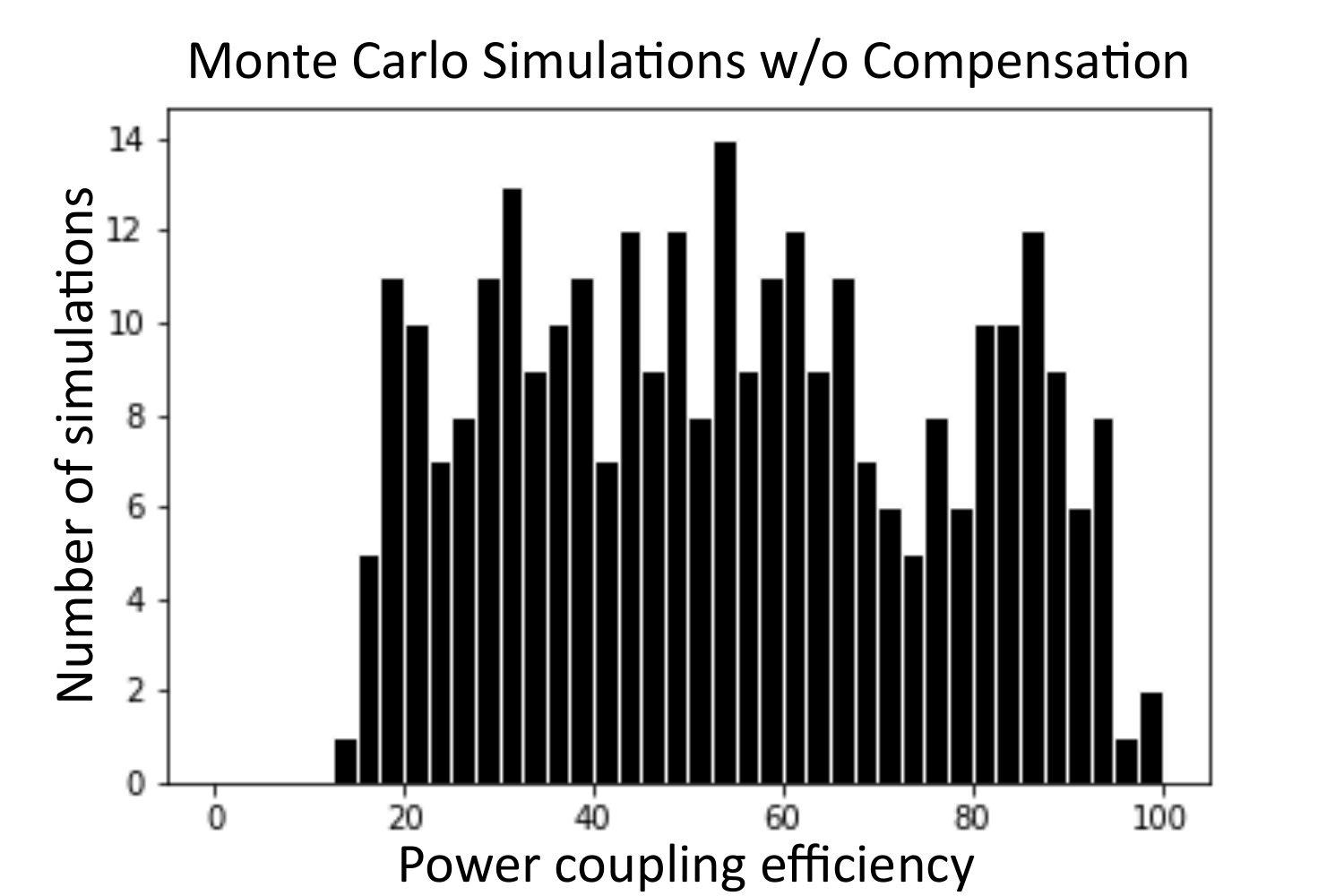}
    \includegraphics[width=8.9cm]{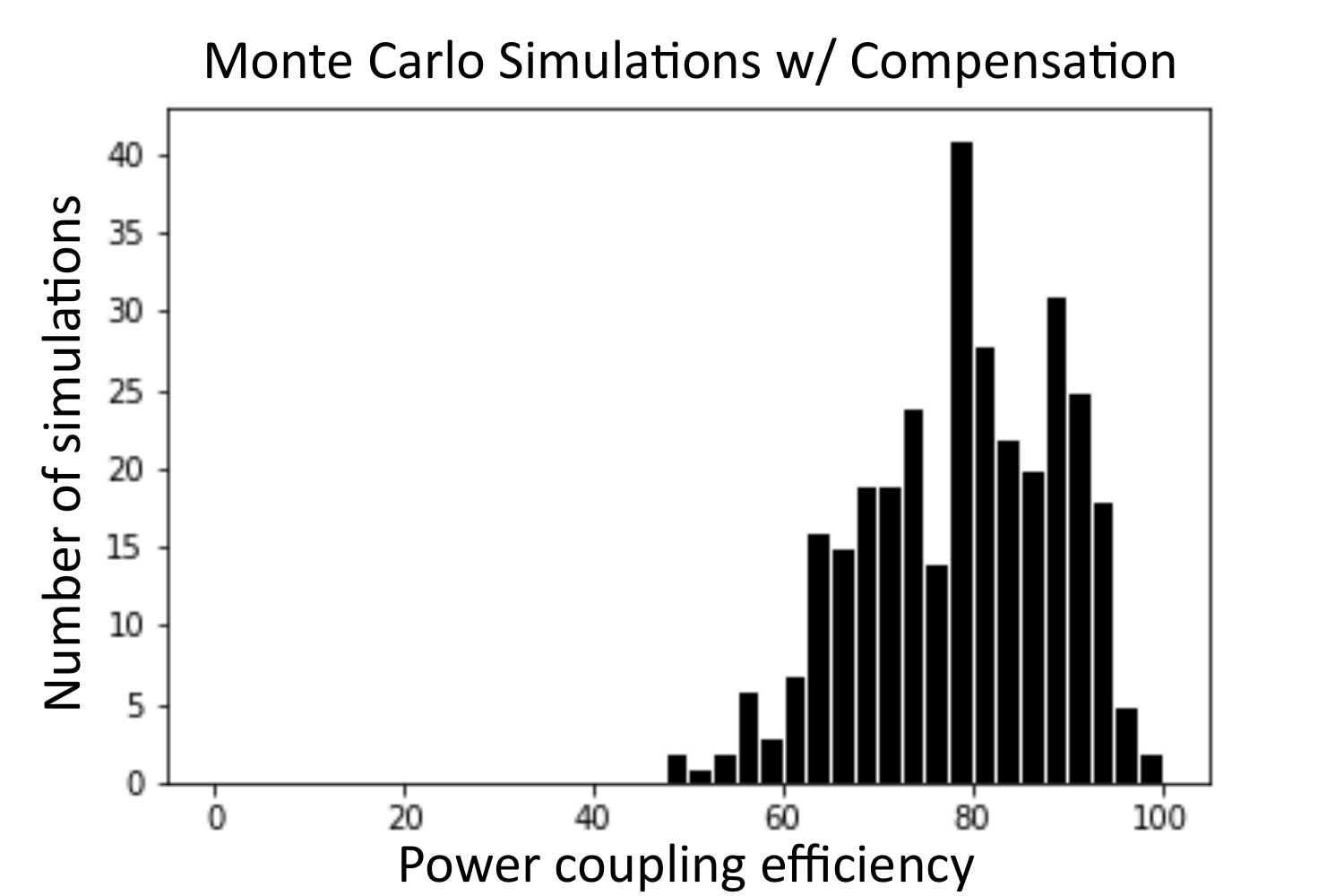}
    \caption{Monte Carlo distributions showing simulation results from our sensitivity analysis. The left figure shows the resulting distribution without compensation of the input beam, and the right figure shows the improvement when compensation is used.}
    \label{sims}
\end{figure*}

Aside from the nominal optical design, we conducted a thorough sensitivity analysis to determine the effects of manufacturing tolerances and alignment errors on the 00 mode power coupling efficiency. The primary tolerances we investigated that are critical to the power coupling efficiency are the focal length, thickness, position, tilting, and decentering of the mode matching lenses and cavity mirrors. This investigation was done with a Monte Carlo analysis whereby misalignments and manufacturing errors were randomized throughout the optical system, and the response of the cavity was then simulated. We first used FINESSE simulations to determine the sensitivity of each parameter individually. In general, we found that the tolerance of the mode matching parameters (focal length, lens thickness and separation) were more manageable than the alignment parameters, which can be seen in Table \ref{Tolerance Parameters Table}. To demonstrate the combined effects of these tolerances, we created a Monte Carlo sensitivity analysis using both Zemax and FINESSE simulations. Zemax has a built-in Monte Carlo tolerancing functionality, and was utilized to perform ray-tracing and Gaussian beam propagation while randomizing each tolerancing parameter with a uniform probability centered around the nominal value. Data obtained from the ray-tracing and Gaussian beam propagation for many randomized Monte Carlo simulations would be recorded. This data was used to calculate the misalignment and mode-mismatch in the beam incident on the cavity, which was then simulated in a FINESSE environment to determine the impact on the 00 mode power coupling efficiency. The mismatches and misalignment in each mode matching lens would result in an incident beam with a Gaussian mode different from the cavity eigenmode, as well as an angle and offset from the ideal cavity optical axis \cite{siegman_1986}. This translates to a lowered coupling efficiency into the cavity 00 mode when simulated with FINESSE. The results of this analysis show that even with tight constraints on the lens alignment ($\le$ 100 $\mu$rad tilt and $\le$ 5 $\mu$m offset) and precision-grade tolerances on all other parameters (Table \ref{Tolerance Parameters Table}), the Monte Carlo distribution we obtained still had an average of 55\% $\pm$ 20\% power coupling efficiency, as shown in Figure \ref{sims} (left).

\begin{table}[h]
    \centering
\begin{tabular}{|m{2.8cm}||m{2.5cm}|m{2.8cm}|}
\hline
 \textbf{Parameter} & \textbf{Tolerance Range} & \textbf{Simulated Effect} \\
\hline \hline
 Lens 1 Surface 2 & 15.979 mm & V = 98.4 \% (+) \\
 radius of curvature & $\pm$ 0.034 mm & V = 98.4 \% (-) \\
\hline
 Lens 1 & 4.45 mm & V = 99.9 \% (+) \\
 thickness & $\pm$ 50 $\mu$m & V = 99.8 \% (-) \\
\hline
 Lens 1 - Lens 2 & 17.26 mm & V = 99.0 \% (+) \\
 Separation & $\pm$ 50 $\mu$m & V = 99.3 \% (-) \\
\hline
 Lens 1 decenter & $\pm$ 5 $\mu$m & V = 72.6 \% ($\pm$) \\
\hline
 Lens 1 tilt & $\pm$ 140 $\mu$rad & V = 93.4 \% ($\pm$) \\
\hline
\end{tabular}
\caption{Simulation results showing the effect of tolerances on different optical parameters in the OTI input stage. The simulated power coupling efficiencies shown in the right column are taken at the upper and lower limits of each parameters tolerance range while leaving all other parameters ideal. Our investigations show that the worst offenders are the tip/tilt and decentering of the mode-matching lenses.}
\label{Tolerance Parameters Table}
\end{table}

Rather than to tighten the tolerance constraints beyond manufacturing feasibility, the most practical way to improve performance was to compensate the lens misalignments with the alignment of the input beam. In practice, this can be done with an adjustable fiber collimator with angular and lateral degrees of freedom. In Zemax, we modeled this by writing a script that interacts with the Monte Carlo analysis where, for each randomized simulation, the alignment of the input beam (output from the collimator) is adjusted to minimize the misalignments in the beam incident on the cavity. In other words, the incident beam can be aligned closer to the ideal cavity optical axis to improve the power coupling to the 00 mode. Tuning the precision and dynamic range of the input beam alignment was also allowed to create a realistic model of the compensation. The resulting Monte Carlo simulations with the compensation of the input beam included produced a distribution with an average of 78\% $\pm$ 10\% power coupling efficiency (see Figure \ref{sims} (right)). Thus, a fiber collimator with precision alignment capabilities is a necessary component to accompany each input stage.

\subsection{Optomechanical Design}
\label{sec:2.4}
The mechanical design of the input stage, as well as the return stage, was fixated on creating a highly thermally stable housing for the optical components. Athermalization of the OTI was primarily done by choosing materials with low coefficients of thermal expansion (CTE) \cite{optical_design_Fischer}. As such, we selected Schott’s Zerodur glass-ceramic (CTE of $10^{-8}-10^{-7} \mathrm{K^{-1}}$) and Invar 36 nickel-iron alloy (CTE of $10^{-6} \mathrm{K^{-1}}$) to form the optical housings, while the optics themselves are made of fused silica with a CTE of around $5 \times 10^{-7} \mathrm{K^{-1}}$. The primary concern for the performance of the OTI is the stability of the cavity mirror surfaces relative to both the housing stages as well as the structure on which the units are mounted, as it is the separation between the cavity mirrors that will be measured in future experiments. The goal is for the cavity displacement noise to be driven by the testing structure on which the cavity is mounted, and not to be driven by the thermal expansion of the optical housings and components. We aim to measure structures which have a $\frac{\mathrm{pm}}{\sqrt{\mathrm{Hz}}}$ displacement noise in the LISA observation band, so the expected thermal expansion of both the optics and the housing components should ideally not be greater than this noise level. Future OTI testing experiments will be in a thermally-shielded vacuum chamber with an expected temperature stability of $10^{-5} \frac{\mathrm{K}}{\sqrt{\mathrm{Hz}}}$ \cite{Armano_2019}. See Table \ref{Optomechanical Design Table} for the expected thermal expansion generated by the optical components and housings assuming slowly varying temperature fluctuations of $10^{-5}$ K.

\begin{table}[h]
    \centering
\begin{tabular}{|m{2.6cm}||m{1.5cm}|m{1.5cm}|m{2.25cm}|}
\hline
 \textbf{Component} & \textbf{Length [mm]} & \textbf{CTE [K$^{-1}$]} & \textbf{Displacement [pm]} \\
\hline \hline
 Zerodur housing & 38 & $10^{-7}$ & 0.038 \\
\hline
 Invar 36 sleeve & 37.5 & $1.3 \times 10^{-6}$ & 0.488 \\
\hline
 Fused silica optics & 4.5-5.6 & $5.8 \times 10^{-7}$ & 0.026-0.033 \\
\hline
\end{tabular}
\caption{Expected thermal expansion due to a $10^{-5}$ K temperature variation in the primary components in the OTI input and return stages. This is only looking at thermal expansion along the optical axis, which is the primary concern for future experiments. All components displace by less than a picometer in the expected thermal environment.}
\label{Optomechanical Design Table}
\end{table}

\begin{figure*}[htpb]
    \centering
    \includegraphics[width=5.9cm]{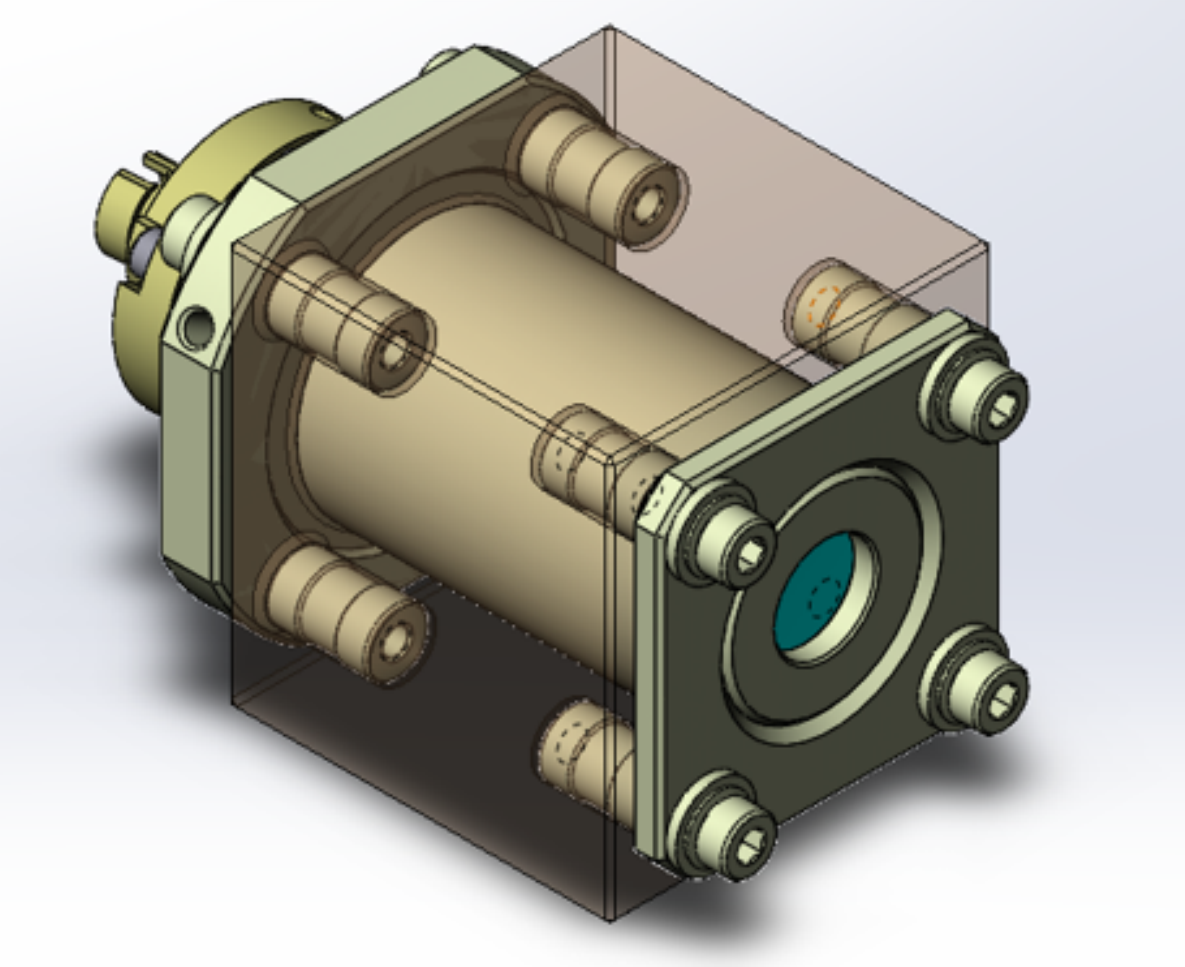}
    \includegraphics[width=8.5cm]{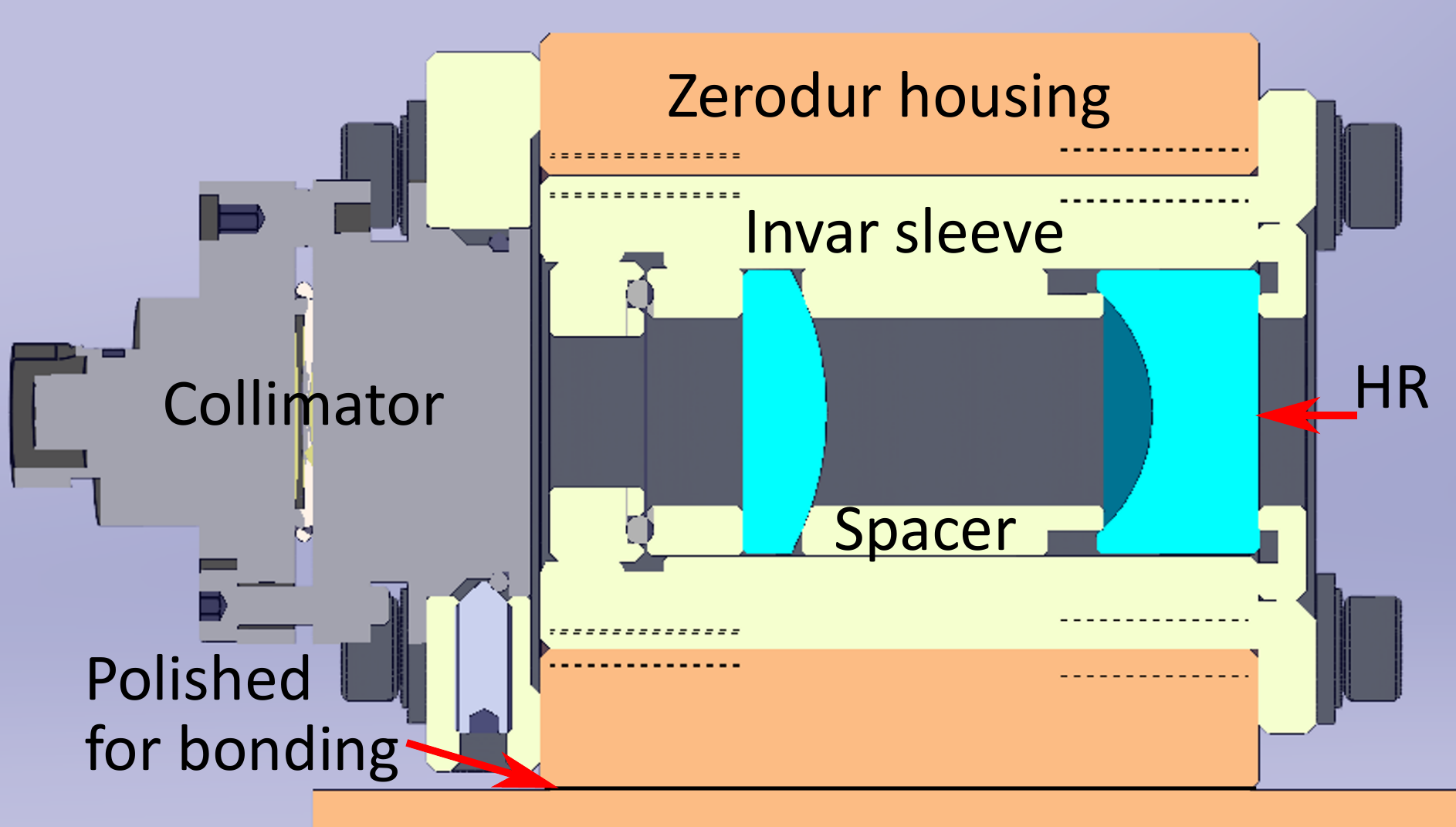}
    \caption{Three-dimensional and cross-sectional diagrams of the OTI input stage. An Invar sleeve, which holds the fused silica components in place, is inserted into a Zerodur housing which can then be bonded/secured to a test structure or telescope. The return stage has a similar design, but only houses the cavity return mirror.}
    \label{Mechdesign}
\end{figure*}

While the optical and mechanical components in the OTI units would nominally allow for a $\frac{\mathrm{pm}}{\sqrt{\mathrm{Hz}}}$ displacement sensitivity, the optomechanical system was still designed to compensate for the thermal expansion of these components. This is facilitated by a cylindrical Invar sleeve (Figure \ref{Mechdesign}) which holds the fused-silica optics and is seated inside of a Zerodur housing polished to $\frac{\lambda}{10}$ ($\lambda$ = 633 nm) on one side for bonding. There are spacer rings placed inside of the metal sleeve that keep a proper separation between the lenses and center them with respect to each other. As shown in Table \ref{Optomechanical Design Table}, Invar 36 has a higher CTE than both Zerodur and fused silica and would reach nearly 0.5 pm displacement under a $10^{-5}$ K temperature variation. For this reason, the mounting of the sleeve to the Zerodur housing is designed to allow the expansion of the metal while maintaining the position of the high-reflectance (HR) cavity mirror surface. Graphical representations of the optomechanical design of the input stage are shown in Figure \ref{Mechdesign}, where the Invar components are highlighted in yellow and the Zerodur components in orange. Assuming the glass components do not expand, the Invar sleeve and spacers will expand toward the fiber collimator into the retaining ring, and the HR cavity mirror surface will remain relatively fixed. Thus, the optomechanical design for the OTI units has been properly athermalized to allow for a $\frac{\mathrm{pm}}{\sqrt{\mathrm{Hz}}}$ displacement noise when bonded or mounted to a LISA telescope or test structure.

\section{Prototype Alignment \& Future Testing}
\label{sec:prototypes}

\subsection{Preliminary Testing}
\label{sec:3.1}

\begin{figure*}[htpb]
    \centering
    \includegraphics[width=7.5cm]{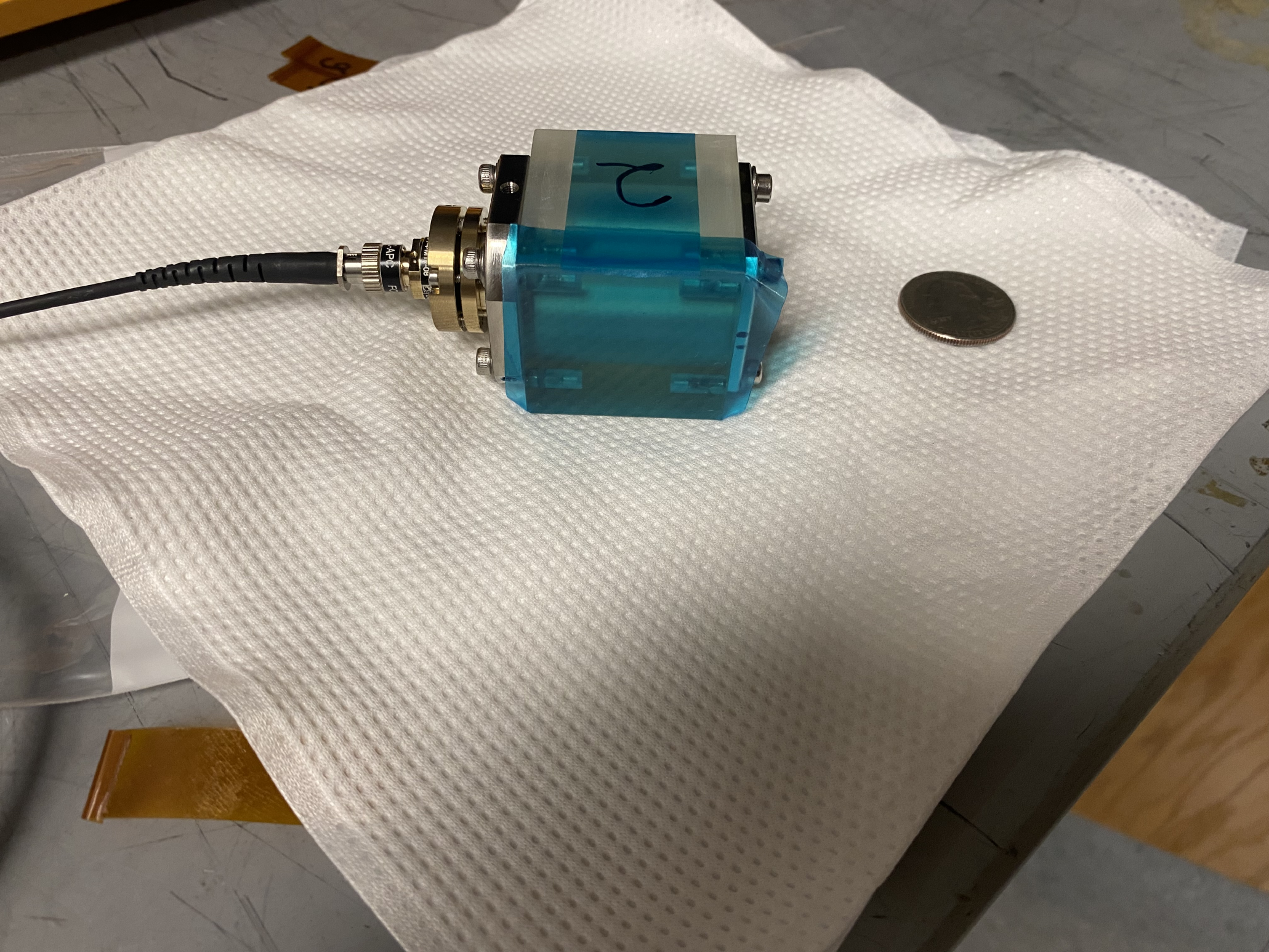}
    \includegraphics[width=7.5cm]{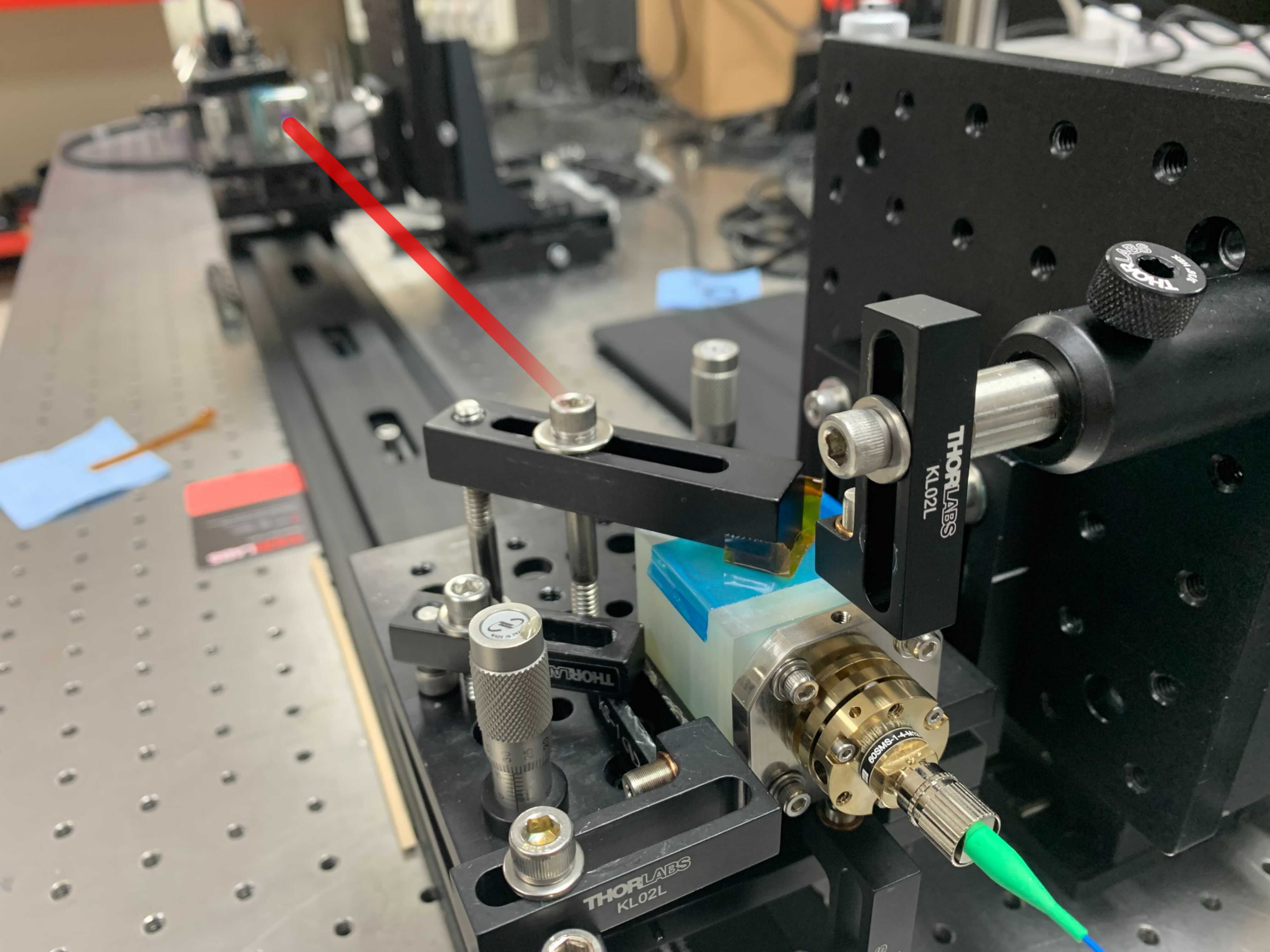}
    \caption{Photos of the first-generation prototypes after fabrication and assembly. The left photo shows an OTI input stage with a quarter for a size reference. The right shows an OTI cavity aligned using commercial translation and rotational stages. The input stage is portrayed in the foreground, while the return stage is out of focus in the background.}
    \label{OTI Ruda Cavity}
\end{figure*}

\begin{figure*}[htpb]
    \centering
    \includegraphics[width=15cm]{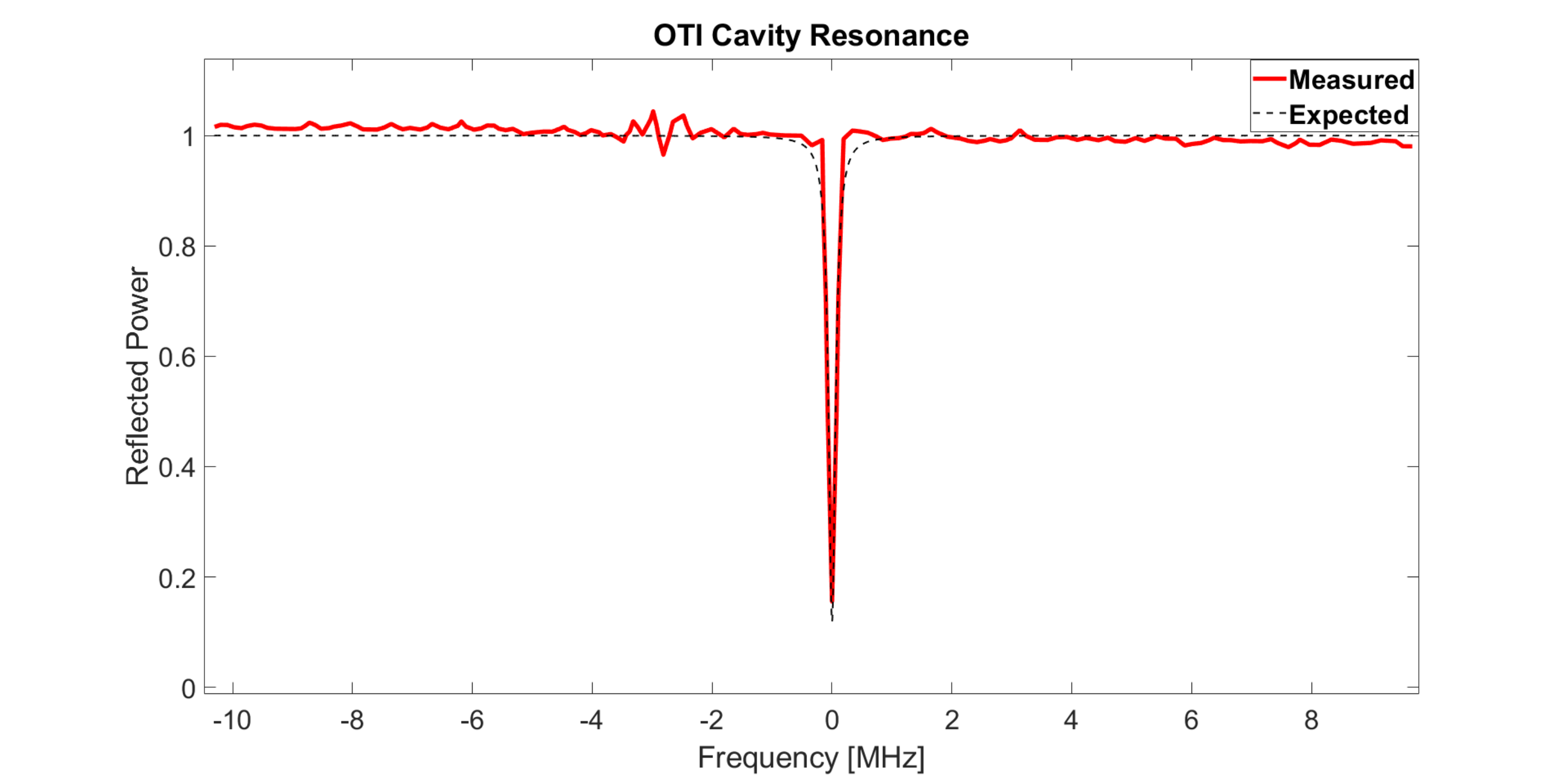}
    \caption{Plot generated from data taken during preliminary testing of the prototype OTI cavities. The red curve shows the reflected power measured by a photodiode while the laser scanned over the cavity resonance. The dashed black curve shows the expected resonance signal based on the average performance from our Monte Carlo analysis.}
    \label{Resonance}
\end{figure*}

We conducted preliminary tests after three first-generation OTI cavity prototypes were fabricated and assembled. There were two key characteristics that we measured to quantify the performance of the prototypes. The first was the fraction of the power in the incident beam that reflected from the HR mirror surface and coupled back into the fiber, which was measured at the output of a fiber circulator to isolate the incident and reflected light. This indicates the internal losses in the OTI input stage and the fiber components along the beam path (mating sleeves, circulator). The second characteristic was the power coupling efficiency measured in the reflected power as the frequency of the laser was swept across the resonance of the fundamental cavity mode. This quantifies the power coupling into the cavity and indicates the overall alignment of the optical system. We started by mounting the fiber collimator to an Invar plate which is attached to the Zerodur housing, as shown in Figure \ref{OTI Ruda Cavity}. The position of the plate could be slightly adjusted to provide lateral translation, and the collimator came equipped with angular adjustment screws. Utilizing these degrees of freedom to optimize the alignment of the incident beam and re-couple the reflected beam, roughly a third of the injected power was measured out of the third port of the circulator. Considering the losses through the fiber components (varying loss at mating sleeves, 0.99 dB loss from port 2 - port 3 in circulator), we found that 70 – 75\% (1.25 – 1.55 dB loss) of optical power injected into the input stage coupled back into the fiber (see Table \ref{Prelim Testing Table}). This loss is most likely due to limitations in the manufacturing and alignment of the optical components in the input stage, as well as the precision with which we could align the fiber collimator. Next, we aligned the prototype cavities by mounting the OTI units on translational and rotational stages (Figure \ref{OTI Ruda Cavity}) to align the cavity mirrors with respect to each other with a 70 cm separation. In all three prototype cavities, we were able to align the system to produce 70-80\% power coupling efficiency in the reflected power when the laser frequency was swept across the fundamental cavity mode as shown in Figure \ref{Resonance}. These results agree with our expectations and predictions based on our simulation model, and the performance between all three prototypes was uniform as shown in Table \ref{Prelim Testing Table}.

\begin{table}[h]
    \centering
\begin{tabular}{|m{3cm}||m{1.5cm}|m{1.5cm}|m{1.5cm}|}
\hline
 \textbf{Optical Power Loss} & \textbf{Unit \#1} & \textbf{Unit \#2} & \textbf{Unit \#3} \\
\hline \hline
 PM mating sleeves & 0.75 dB & 1.02 dB & 1.14 dB \\
\hline
 PM fiber circulator & 0.99 dB & 0.99 dB & 0.99 dB \\
\hline
 OTI Input Stage & 1.25 dB (75 \%) & 1.49 dB (71 \%) & 1.37 dB (73 \%) \\
\hline \hline
 \textbf{Cavity Alignment} & \textbf{Unit \#1} & \textbf{Unit \#2} & \textbf{Unit \#3} \\
\hline
 $V = \frac{P_{max}-P_{min}}{P_{max}+P_{min}}$ & 75.7 \% & 73.1 \% & 79.1 \% \\
\hline
\end{tabular}
\caption{Results of characterizing optical power losses through the fiber components and OTI input stages, as well as cavity 00 mode coupling, during preliminary testing. The total power loss was measured as the ratio between the powers measured at port-2 and port-3 of the fiber circulator, and the internal loss through each OTI input stage was calculated using the measured loss through each fiber component.}
\label{Prelim Testing Table}
\end{table}

\subsection{Methodologies for Future Testing}
\label{sec:3.2}

\begin{figure*}[htpb]
    \centering
    \includegraphics[width=15cm]{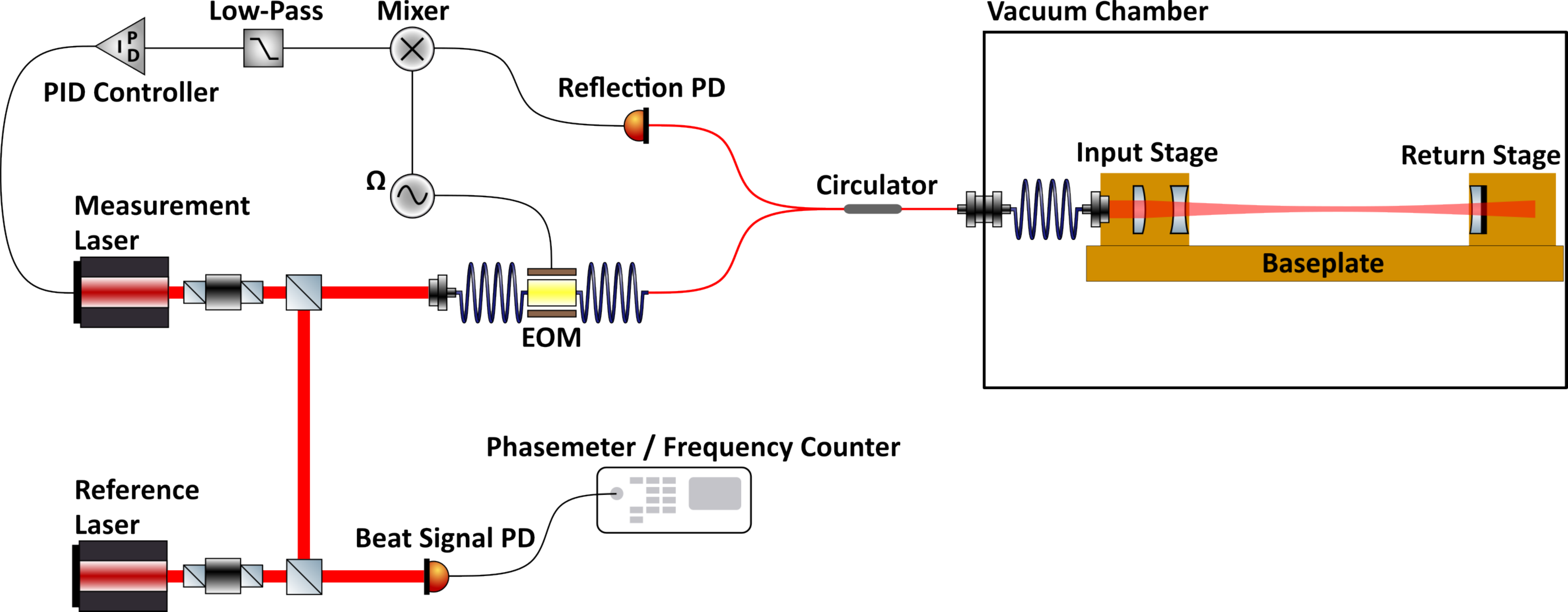}
    \caption{Schematic for an experimental test bed to verify the desired displacement sensitivity of the OTI prototypes. The input and return stages are mounted and aligned on a baseplate made of low-expansive material (left), placed in a vacuum environment, and the cavity is injected with a 1064 nm laser field that is used for a PDH-based readout scheme (right).}
    \label{pm Testing}
\end{figure*}

The prototypes must be further tested and verified to be capable of the necessary $\frac{\mathrm{pm}}{\sqrt{\mathrm{Hz}}}$ displacement sensitivity in the mHz observation bandwidth, so a proper test bed must be made for the prototypes and the readout schemes that will accompany them. The methodology to test the OTI system is primarily based on the Pound-Drever-Hall (PDH) frequency locking technique. The basic setup for this, depicted in Figure \ref{pm Testing}, consists of a fiber-coupled 1064 nm laser source, an electro-optic modulator (EOM), a fiber circulator, high-bandwidth photodetectors, and the OTI cavity along with other supporting optical and electronic components. The fiber-based EOM induces a sinusoidal phase modulation in the injected beam when driven at an RF modulation frequency $\Omega$. The alignment between the applied voltage across the EOM crystal and the polarization of the incident field is crucial to the efficiency of the modulation, which is why polarization-maintaining fibers are employed throughout the system. The fiber circulator isolates the field incident on the cavity from the field reflected from the cavity, which is measured by a photodetector. The resulting signal produced by the reflected field is demodulated at the modulation frequency $\Omega$ to generate the well-established PDH error signal upon scanning the laser frequency over the cavity resonance \cite{black_2001}. The error signal is used to lock the laser to the cavity resonance via a PID controller feedback loop, thereby coupling the laser frequency fluctuations $(\delta f)$ to the OTI cavity length fluctuations $(\delta L)$ through the relation $\frac{\delta f}{f} = \frac{\delta L}{L}$. Thus, monitoring the laser frequency noise effectively monitors the displacement noise in the structure to which the OTI cavity is mounted. This is done by recording the beat frequency between the measurement laser and a reference laser, as shown in Figure \ref{pm Testing}. The reference laser serves as a stable frequency such that the noise in the beat signal is dominated by the measurement laser, which can be achieved, for example, by stabilizing the reference laser to a separate ultra-stable cavity.

\begin{figure*}[htpb]
    \centering
    \includegraphics[width=15cm]{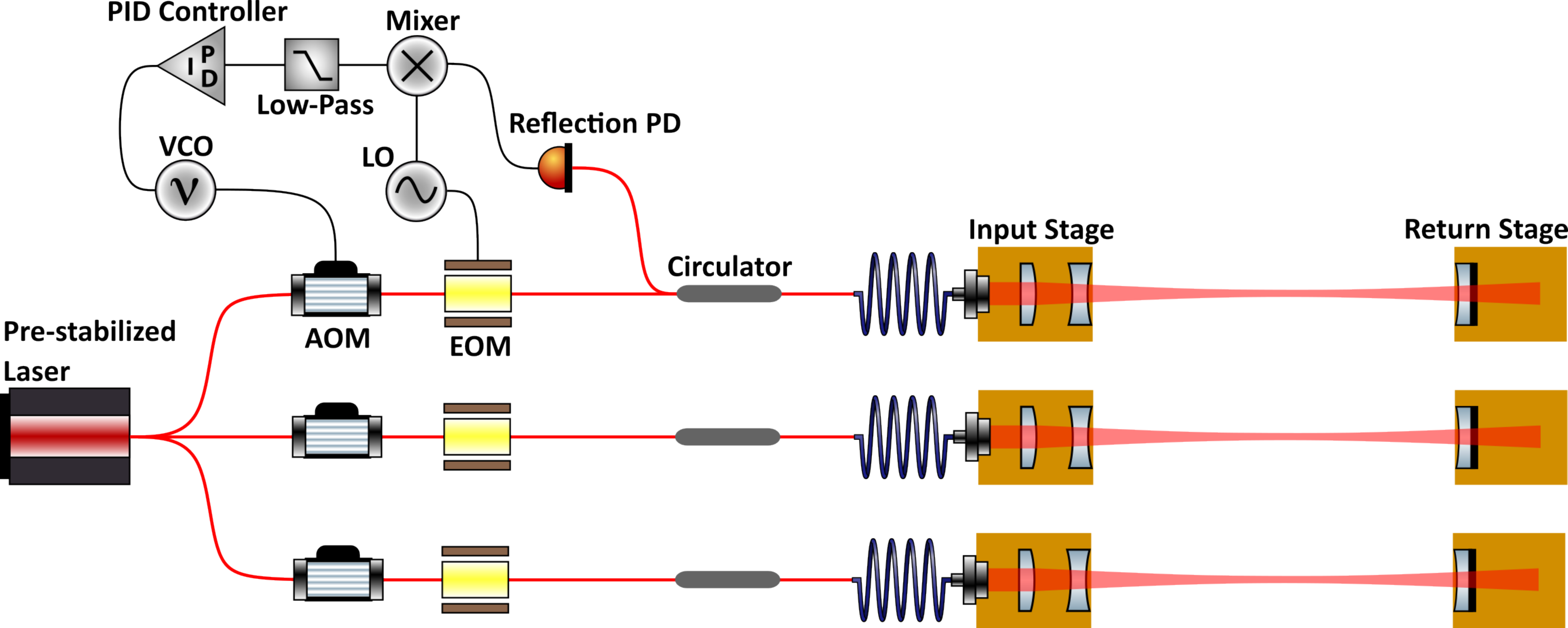}
    \caption{Diagram of a possible readout scheme to measure the displacement noise in multiple OTI cavities simultaneously. The feedback control signal for each cavity is sent into a voltage-controlled oscillator (VCO) which drives the frequency shift induced by the AOM.}
    \label{Parallel Testing}
\end{figure*}

While the standard PDH technique is sufficient to monitor the displacement noise in an OTI cavity, it can become cumbersome and costly to use this method with multiple cavities simultaneously as depicted in Figure \ref{OTI_inkscape}. To further demonstrate the working principle behind the OTI, we aim to design a readout scheme that can measure multiple OTI cavities simultaneously using only a single laser source. One possible solution is to utilize a combination of an acousto-optic modulator (AOM) and an EOM to accompany each cavity, as shown in Figure \ref{Parallel Testing}. In this system, the AOM is used to shift the nominal laser frequency and to tune that frequency shift. The EOM is used to modulate the phase of the beam to create the standard PDH signal in the beam reflected from the cavity. However, instead of feeding the control signal back into the laser source, the signal can be fed back into the AOM driver to tune the frequency shift and keep the carrier frequency locked to the cavity resonance. In principle, this allows a single laser source to operate multiple OTI cavities simultaneously. This laser source, however, must be pre-stabilized similarly to the reference laser in Figure \ref{pm Testing} such that the frequency noise does not interfere with the measurements. In this system, the frequency driving each AOM will shift in proportion to the cavity resonance and thus can be measured to determine the displacement noise in each cavity. The standard PDH methodology, as well as parallel readout schemes such as this, will be investigated in future testing of the OTI prototypes.

\section{Conclusion}
\label{sec:conc}
The optical truss interferometer is designed to be a robust optomechanical system composed of compact, modular units that can be integrated with the LISA telescope prototypes for ground testing, as well as a risk-mitigation plan for flight units. We have designed, procured, and performed preliminary tests on three first-generation OTI prototype cavities. The optical design of the OTI cavity minimizes sensitivity to misalignments and manufacturing errors, while the mechanical design prioritizes low thermal expansion to minimize displacement noise in the cavity baseline due to the OTI units themselves. We have designed a 70 cm symmetric cavity with a finesse of  $\gtrapprox$ 1600, and an input stage that integrates a fiber-coupler, mode matching lenses, and cavity input mirror into a single, handheld unit less than 5 cm in length. We were able to align these cavities after assembly and achieved around 75\% power coupling efficiency in the 00 mode for all three units, confirming the predictions from our simulation models. These prototypes show promising results from the initial alignment and characterization of the cavities, and we are ready for the next stage of testing where we will verify a $\frac{\mathrm{pm}}{\sqrt{\mathrm{Hz}}}$ displacement sensitivity using various PDH-based methodologies to fully demonstrate the working principle behind the optical truss interferometer.
\newpage

\section*{Funding}
The authors gratefully acknowledge the financial support from the NASA Goddard Space Flight Center (Grant 80NSSC22K0675) and the University of Florida (Sub-Award SUB00003369).

\section*{Acknowledgments}
The authors acknowledge Jose Sanjuan Munoz, Ph.D. and Xiangyu Guo, Ph.D. of the Laboratory of Space Systems and Optomechanics (LASSO) at Texas A\&M University for editing the contents of this manuscript.

\section*{Disclosures}
The authors declare no conflicts of interest.

\section*{Data Availability}
Data underlying the results presented in this paper are not publicly available at this time but may be obtained from the authors upon reasonable request.

\bibliography{Bibliography}

\end{document}